\documentclass[aps]{revtex4}
\usepackage{graphicx,subfigure,color}
\begin{document}
\title{Memory effects in tumbling nematics of 8CB liquid crystals}
\author{Balaji Yendeti,  Ashok Vudaygiri\footnote{corresponding author: email avsp@uohyd.ernet.in}}
\affiliation{School of Physics, University of Hyderabad, Hyderabad, 500046 {\bf India}}
\begin{abstract}
Particle tracking with a 0.98$\mu$m silica sphere is used in determining precessional motion of nematic director in nematic phase of 8CB liquid crystals, as it probes those oriented structures which are of the same wavelength as of the sphere size. Velocity auto correlation(VACF) is used in determining those structures in both parallel and perpendicular orientations to the neamatic director. Further, a generic approach by considering the time dependent harmonic oscillator motions is used to analyze the VACF distribution function. This approach leads to observe a transition in the structures of nematic phase that are comparable to transformations from underdamped harmonic oscillator motion to critically damped motion. Also, we measured the microstructural properties and calculated micromechanical properties.  The experimental analysis approach used here for 8CB liquid crystals helps to understand and characterize the general dynamic behavior of complex fluids. With this analysis, `dynamics of complex fluids'  becomes no more `complex'. 
\end{abstract}  
\maketitle
\section{introduction}
In nature, many of the things we observe are either time dependent or frequency dependent\cite{P.M.Chaikin}.  Dynamic correlations are described by time-dependent generalizations of correlation functions which contain information about the nature of dynamical modes. The dynamical properties of condensed matter systems are mostly dominated by harmonic oscillator-like modes.  Correlation functions contain information about the frequency and damping of these modes. Velocity auto correlation function(VACF) can be used for the quantitative description of dynamics of molecular motion in fluids. The VACF denoted by $C_{v}(t)$ is the average of the initial velocity $V(0)$ of a particle multiplied with its  velocity $V(t)$ at a later time  t,  $C_{v}(t) = \langle V(t)V(0) \rangle$ \cite{M. H. J. Hagen}. Physically, it represents, temporal correlations between the random walkers and the background in a disordered medium.  Any energy in the harmonic oscillator will tend to flow irreversibly into many modes of the fluid. This irreversible flow of energy into incoherent degrees of freedom, which is called as dissipation, is reflected in the sign of the viscous force which is decay of $x(t)$ with time.   $C_{v}(t)$ is a fundamental function,  which is useful to obtain the systems physical properties like diffusion constant$(D)$ from Kubo$'$s relations  as\cite{R. Kubo-1},\cite{R. Kubo-2}
\begin{equation}
D = \lim_{t \rightarrow \infty} \frac{1}{2t}\langle x^{2}(t)\rangle = \int_{0}^{\infty}C_{v}(t)dt
\label{eq-1}
\end{equation}
In nematic liquid crystals, the average orientation of molecules is determined by the nematic director $\hat n (r,t) $. The orientational order in a nematic liquid crystal effects the Brownian motion of spherical particles of micro meter size which leads to anisotropic diffusion\cite{J.C.Loudet-1}. In standard liquid crystals, like 5CB, The director field ${{\hat n} \left( r,t \right) } $ is coupled to the velocity field V(r,t) of the nematic. This coupling of the nematic director with its fluctuations  and velocity field are perturbed by the spherical particle in particle tracking experiments. Hence, the time averaged distribution of ${\hat n  \left( r,t\right) } $ under flow can be characterized by \textit{in situ} measurements   of tracking Brownian motin of colloidal particle. 
 
   In 8CB liquid crystals, along with phase transitions, there exists several regimes in nematic phase \cite{C. R. Safinya},\cite{C R Safinya-1}\cite{K. Negita},\cite{K. Negita-1}. Various structures in different regimes are mentioned with different notations. Even though, these structures were defined in presence of shear,  in this paper, we follow the same notations as prescribed by K.Negita et.al.,\cite{K. Negita-1} in representing structural regimes.
    In these structurqal regimes, tumbling motion of the director $(\hat{\textbf{n}})$  is expected to occur\cite{D.F.Gu},\cite{Dennis J. Ternet},\cite{P.T.Mather}.  As the director precesses about neutral or vortex direction with angular frequency $\omega_{0}$, the equation of motion is described by
 \begin{equation}
  \frac {n_{y}(t)^{2}}{{n^{2}_{y0}}} + \frac{n_{z}(t)^{2}}{n^{2}_{z0}} = 1
  \label{eq-2}
 \end{equation}
 is an equation of ellipse. Here $n_{y}(t) = n_{y0}$ cos$(\omega_{0}t)$ and $n_{z}(t)= n_{z0}$ sin$(\omega_{0}t) $ are components of the director $ n(t) = (n_{x}(t),n_{y}(t),n_{z}(t))$ precessing around the x-axis with an angular frequency of 
\begin{equation}
\omega_{0} = \left[ \frac{\dot {\gamma}^{2}(-\alpha_{2}\alpha_{3}^{R})}{\gamma^{2}_{1}}\right] ^{\frac{1}{2}} 
 \label{frequency}
\end{equation}

Where $\dot{\gamma}$ is the shear rate, $\alpha_{3}^{R} $ is the renormalized viscosity coefficient of $\alpha_{3} $ including the effect of the $S_{A}$ fluctuations, and $\gamma_{1} = \alpha_{3}-\alpha_{2}$.  Depending on the relative strength of the director components, steady structures that are observed:
			1)$a_{m}$ strucure: anisotropic precession with larger amplitude in the y-direction with $(n_{y0}>n_{z0})$.
			  2)$a_{s}$ structure: isotropic precession with equal amplitude in the y and z direction $(n_{y0}= n_{z0}) $.
			  3)$a(b)$ structure: anisotropic precession with larger amplitude in the z direction $(n_{y0}< n_{z0}) $.
			  4)$a_{c}$ structure: anisotropic precession with its  larger amplitude of motion deflected along the z-axis $(n_{y0}<< n_{z0}) $
			  and in the $S_{A}$ phase, the formation of the layer structure makes the precessional motion impossible.  We also discussed about these structural regimes in 8CB liquid crystals in our previous articles\cite{S. Dhara}, \cite{Balaji}.
			  \begin{figure}
\includegraphics[scale=0.3]{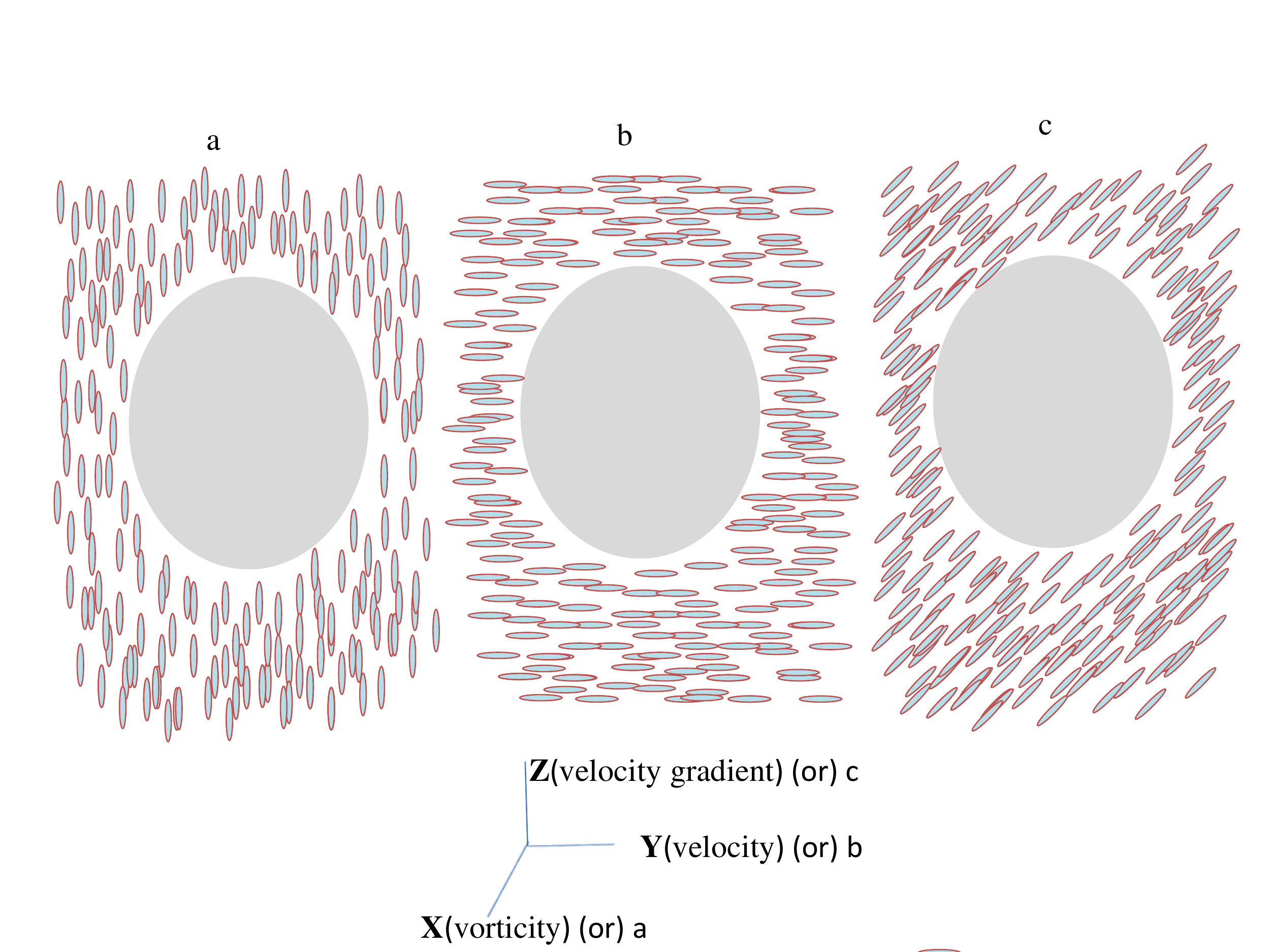}
\caption{a,b,c structures in nematic phase of 8CB liquid crystals represent vorticity(x), velocity(y), and velocity gradient(z) directions.}
\label{8CB picture}
\end{figure}
			  
			  In 8CB liquid crystals  X-ray scattering studies \cite{C. R. Safinya}, discovered that, in the nematic phase, the time averaged distribution of $(\hat{\textbf{n}})$ under flow reveal a sequence of regimes whose occurence results from the interplay between the viscous frictional and flow induced fluctuation forces.  This is an $\textit{in situ}$ X-ray diffraction measurement. Recently, anomalous diffusion, sub and super diffusive regimes and director$(\hat{n})$ deformations around a sphere in standard liquid crystals like 5CB were described by analyzing the $C_{v}$ from thermal motion of colloidal particle\cite{T. Turiv}.  In this paper, we rediscoverd the non-equilibrium flow conditions of the nematic$(N)$ to smectic A$(Sm-A)$ transition in 8CB liquid crystals by analyzing the Brownian motion of colloidal particle. VACF $(C_{v})$, is the rate of change in diffusion that represents the amplitude $r(t)$ of the deformed structures when diameter of the probing particle is nearly equal to the wavelength of the deformed structures. So, $C_{v}$ is used in determining the time averaged distribution of the director $(\hat{\textbf{n}})$  under flow to reveal a sequence of structures in both parallel and perpendicular directions to the director $(\hat{\textbf{n}})$ orientations. These amplitudes and periodicity of the structures are compared with measurements by X-ray diffraction and shear rheology experiments. Further, a generic approach by fitting the $(C_{v})$ with time dependent harmonic oscillations under different conditions is followed.   This approach reveals the kind of structural transitions, micro structural properties like relaxation time $(\tau)$, frequency of oscillations of the director ($\omega$) and micro mechanical properties like elastic constant $(K))$. Also, we measure the quality factor $(Q)$ of  precessional motion of $(\hat{\textbf{n}})$, which represents the stability of fluctuations.  Since, these precessional motions $(\hat{\textbf{n}})$ are known to be anisotropic, in nematic phase of 8CB liquid crystals, it is important to study microstructural and micromechanical properties in both parallel and perpendicular directions to the director orientation.

\section{The Experiment}
For a specific molecular alignment in the 8CB liquid crystals, the colloidal silica spheres of 0.98$\mu m$ in size are coated with a monolayer of Octadecyldimethy(3-trimethoxysilylpropyl)ammonium chloride(DMOAP,sigma aldrich)\cite{V.S.R.Jampani}. This monolayer ensures the strong perpendicular surface anchoring of the nematic liquid crystals with DMOAP coated surfaces\cite{M. Skarabot}. These coated silica spheres are added to the 8CB liquid crystals and sonicated for the dispersion of these beads into liquid crystals. The uniform dispersion of  micro spheres in a planar cell which is coated with AL-1254 on the glass slides and rubbed in one direction, are temperature controlled with a temperature controller(INSTEC) whose resolution is upto $\pm 0.01^{\circ} C $.  In this planar aligned liquid crystal cell, the molecules get arranged in the (x,y) plane and aligned parallel to the y-direciton. That means, the axis of symmetry of the liquid crystals molecules is along y-axis as shown in fig.\ref{8CB picture}.This sample was observed under an inverted polarizing microscope(Nikon Eclipse,T\textit{i} - U) with a microscope objective lens of 100X magnification.   Video microscopy is used in the single particle tracking of an isolated colloidal micro sphere to measure its Brownian motion. The position and Brownian motion of the colloidal particle is measured with the help of an appropriate computer program. The total experimental instrumentation used here are very similar to those mentioned in our previous paper\cite{S. Dhara}. The Brownian motion of an isolated micro sphere is video recorded for about 5 minutes and analyzed. Measurements are repeated with in the same and different cells. Transition characteristics of 8CB liquid crystals are mentioned in our previous work\cite{Balaji}.


\begin{figure}[h]
\end{figure}

\section{Results and discussion}
		In previous experimental measurements \cite{C. R. Safinya, D.F.Gu} the structural regimes in 8CB liquid crystals are analyzed using ELP theory and critical nematodynamics, where shear stress is applied on the material to understand the dissipative nature in various structural regimes and hence the precessional motions of $(\hat{\textbf{n}})$. But, under zero shear conditions, considering ELP theory may not be a appropriate approach. For example, in determining the angular frequency of precessional motion of $(\hat{\textbf{n}})$, equation \ref{frequency} contains shear rate $\dot {\gamma}^{2}$, so, under no shear conditions, structures in 8CB liquid crystals can not be analyzed. 
		
		 In any disordered or complex medium, when the colloidal probe is of the same diameter as of the wavelength of disordered structures, colloidal particle gets navigated through disordered structures.  Hence, single particle tracking technique is used in determining the precessional motion of $(\hat{\textbf{n}})$ and is a correct and simple approach under no shear limits.  	According to Safinya et.al., in 8CB liquid crystals, as the temperature is reduced, for transitions from $N$ to $Sm-A$, the interplay between the viscous frictional forces and the flow induced fluctuation forces on the nematic director leads to a series of structural regimes\cite{C. R. Safinya}. 
			 In the isotropic phase, VACF of 8CB liquid crystals have shown normal diffusive behavior and when the temperature is reduced to just below N-I, at $\Delta T_{N-I} = -0.5^{0}C $, 
			   these structures are flow aligning nematogens, similar to nematic regime of 5CB liquid crystals. In such a case, the coupling of the sphere's displacements to the director field and its fluctuations leads to anomalous subdiffusion. This subdiffusion is defined as fluctuative displacement of the sphere, that temporarily increases elastic energy density  say, on to the left side to its motion and decreases it on the right side. This difference in elastic energy density creates a restoring elastic force $\textbf{F}_{sub}$ which slows down the diffusion.
			 \begin{equation}
			 \textbf{F}_{sub} \approx -K\frac{\Delta\textbf{x}}{d}
			 \label{eq-3}
			 \end{equation}
			 
\begin{figure}
\includegraphics[scale=0.35]{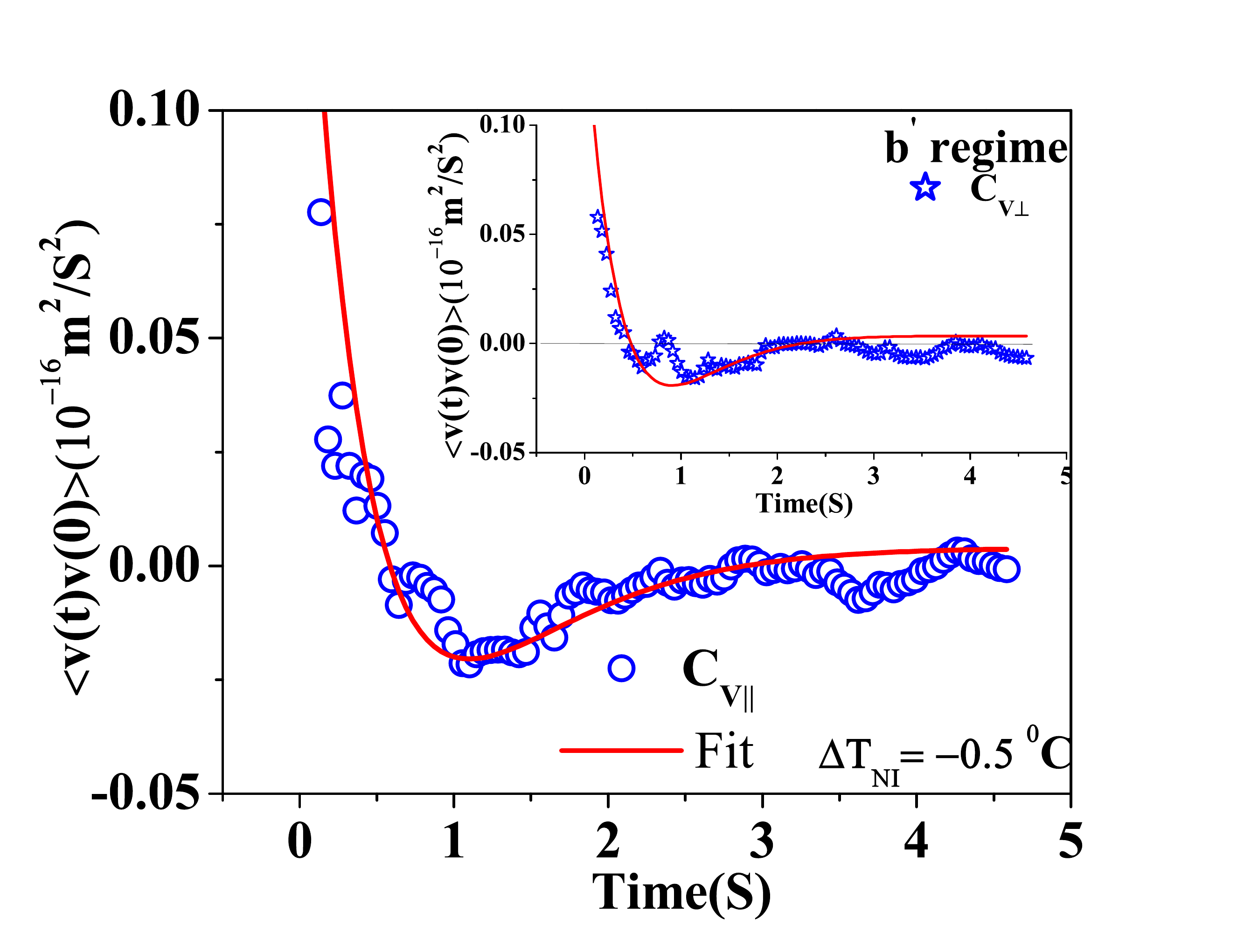}

\caption{At $ \Delta T_{N-I} = -0.5^{0}C $, VACF vs time in $b^{'}$ regime, $C_{v\|}$ is shown by open circles(blue color) and $C_{v\bot}$ is shown by star(blue color)and solid line(red color) is fit.}
\label{figure 2(a)}

\end{figure}

\begin{figure}
\includegraphics[scale=0.35]{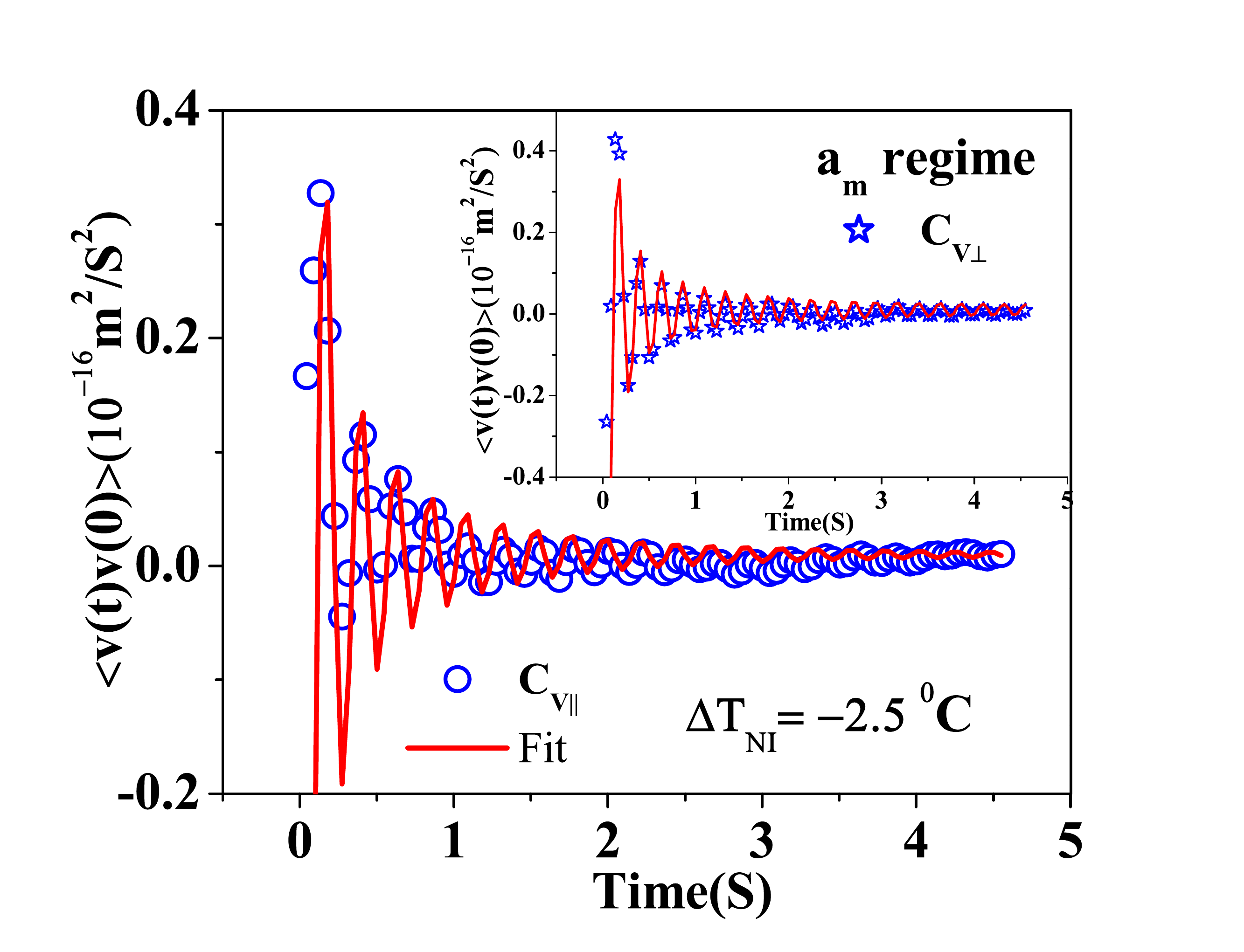}
\caption{At $ \Delta T_{N-I} = -2.5^{0}C $,  VACF vs time in $a_{m}$ regime, $C_{v\|}$ is shown by open circles(blue color) and $C_{v\bot}$ is shown by star(blue color)and solid line(red color) is fit.}
\label{figure 2(b)}
\end{figure}

\begin{figure}
\includegraphics[scale=0.35]{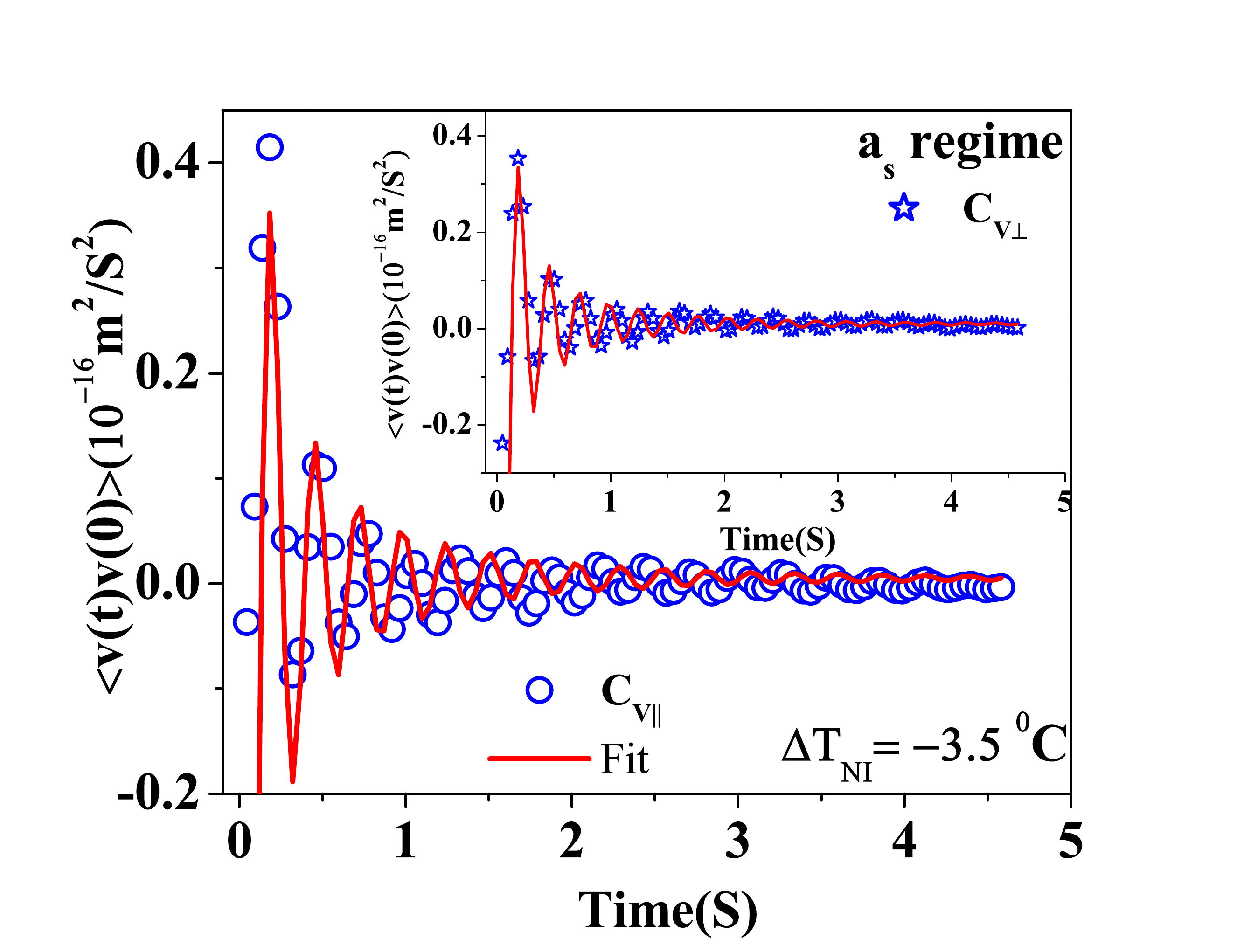}
\caption{At $ \Delta T_{N-I} = -3.5^{0}C $,  VACF vs time in $a_{s}$ regime, $C_{v\|}$ is shown by open circles(blue color) and $C_{v\bot}$ is shown by star(blue color)and solid line(red color) is fit.}
\label{figure 2(c)}
\end{figure}

\begin{figure}
\includegraphics[scale=0.35]{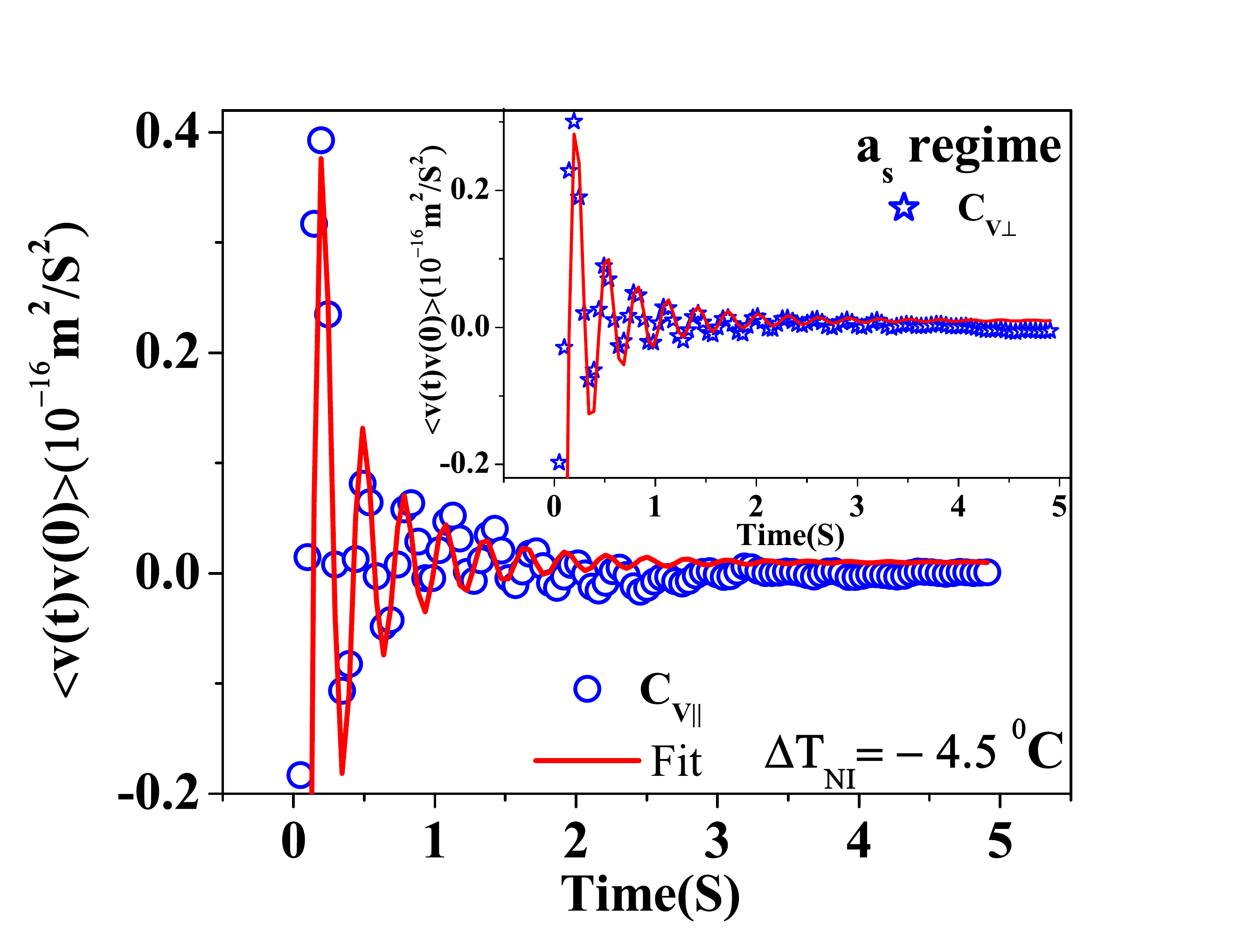}
\caption{At $ \Delta T_{N-I} = -4.5^{0}C $,  VACF vs time in $a_{s}$ regime, $C_{v\|}$ is shown by open circles(blue color) and $C_{v\bot}$ is shown by star(blue color) and solid line(red color) is fit.}
\label{figure 2(d)}
\end{figure}

			 Where $\Delta\textbf{x}$ is the change in displacement from left side to right side.  As shown in figure \ref{figure 2(a)}, values of $C_{v}$ are negative representing sub diffusive nature\cite{T. Turiv}.  But, curve fitting for these VACF gave positive values for the amplitudes in both parallel and perpendicular directions to the director. This represents, the direction of the flow alignment of the director being with its direction near the flow direction(y-axis).  From, $b^{'}$ to $a-b$ transition point, the amplitude of the VACF curves have changed from positive to negative value. 
			 Below $a-b$ regime, because of presmectic translations, there arises few layer formations in this regime. These layers gets rearranged and hence there exists compression on to the central portion of the liquid crystals. Details of the physical origin of the structural regimes in 8CB liquid crystals under zero shear conditions was explained in our previous work\cite{Balaji}.  
			  VACF for structures below $a-b$ regime, were presented as shown in figures \ref{figure 2(a)}, \ref{figure 2(b)}, \ref{figure 2(c)}, \ref{figure 2(d)}. These figures resemble the tumbling nematic director behavior in the apparent viscosity vs shear strain of shear rheology experiments \cite{D.F.Gu}. From Kubo's relation\ref{eq-1}, both memory function $C_{v}$ and viscosity are related, Hence the resemblance. In case of bulk rheolgy experiments, within apparent viscosity vs shear strain,when $\alpha_{3}$ begins to diverge, there is a rapid increase in the oscillation amplitude with the decrease in temperature, and near $T_{Sm-A}$ regime, the damping of  oscillation amplitude occurs rapidly.   Similar characteristics were observed for $ C_{v\|}$.  Where as, $C_{v\bot}$  decreases continuously  in the values of oscillation amplitudes. This damping in oscillation amplitude may be due to several factors like out of plane motion of the director, the suppressive effect of the increasing elastic torque, and the appearance of a poly domain structure. But, clearly the faster decay in oscillation amplitudes near $T_{SmA-N}$ can be attributed to the out of plane motion of the director\cite{D.F.Gu}. Also, as the colloidal particle passes through these structures, it dissipates energy, hence it can be observed as damping in oscillation amplitude.
			 
\begin{figure}
\includegraphics[scale=0.35]{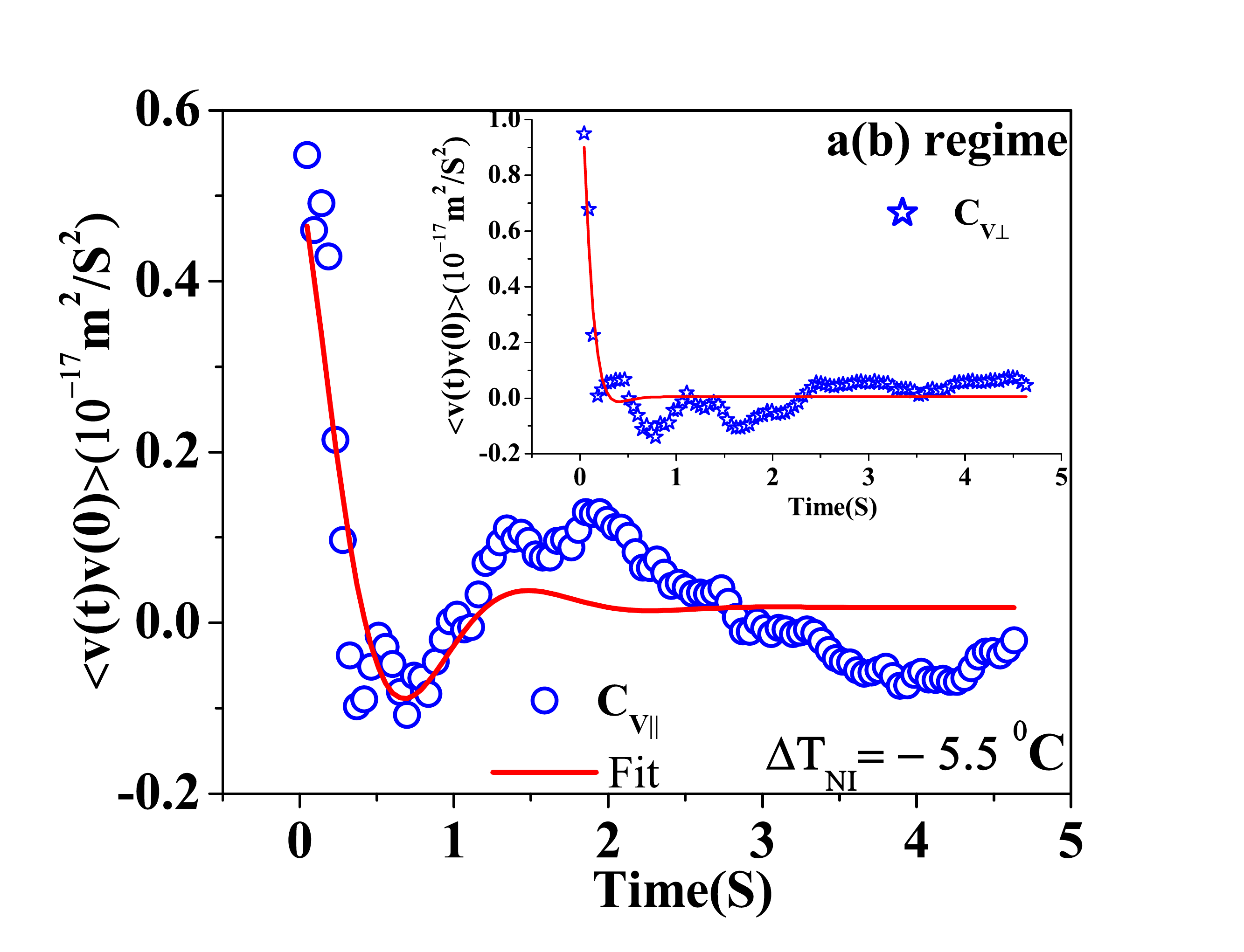}
\caption{At $ \Delta T_{N-I} = -5.5^{0}C $,  VACF vs time in $a(b)$ regime, $C_{v\|}$ is shown by open circles(blue color) and $C_{v\bot}$ is shown by star(blue color) and solid line(red color) is fit.}
\label{figure 2(e)}
\end{figure}
\begin{figure}
\includegraphics[scale=0.35]{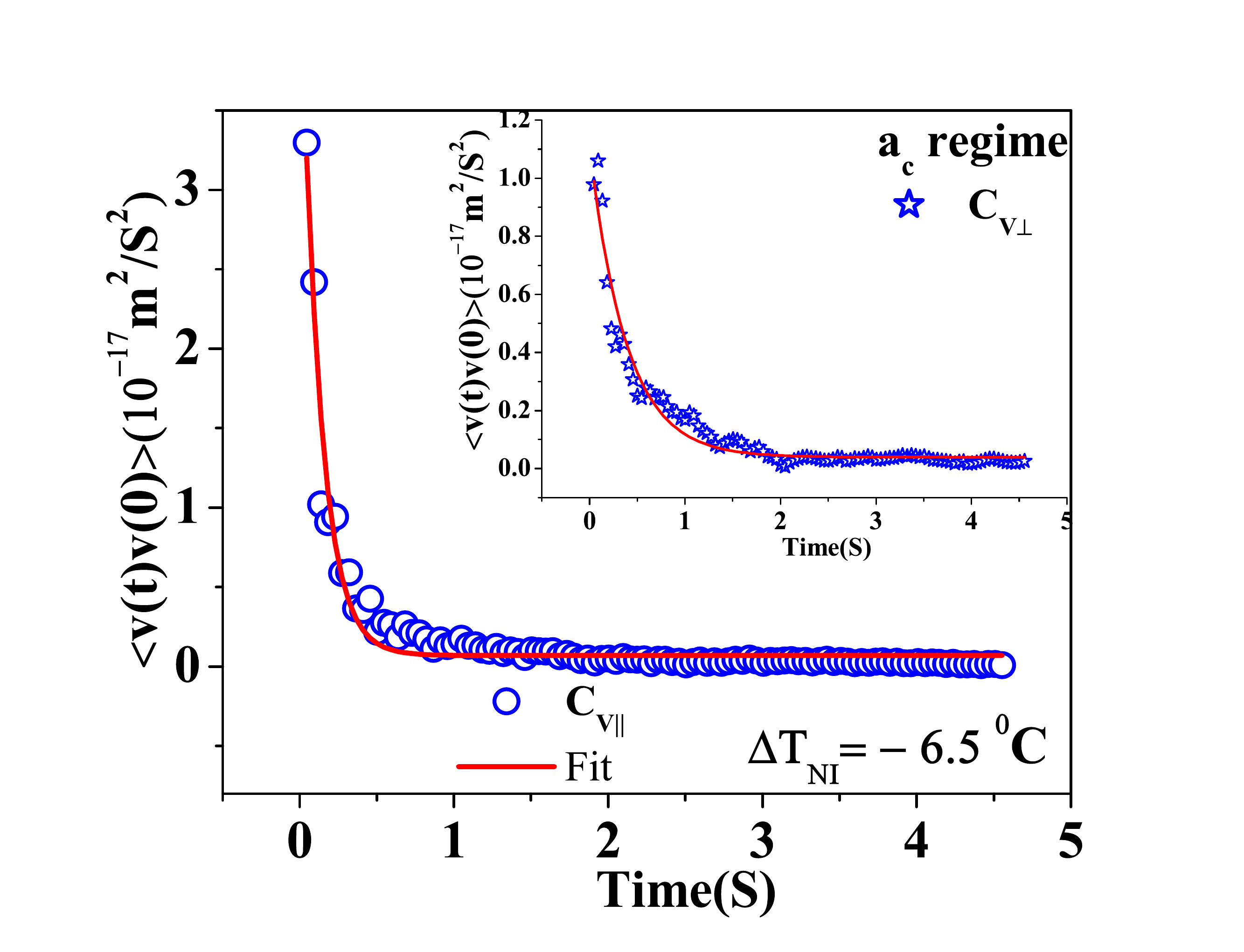}
\caption{At $ \Delta T_{N-I} = -2.5^{0}C $,  VACF vs time in $a_{c}$ regime, $C_{v\|}$ is shown by open circles(blue color) and $C_{v\bot}$ is shown by star(blue color) and solid line(red color) is fit. }
\label{figure 2(f)}
\end{figure}

			      Both $(C_{v\|})$ and $(C_{v\bot})$  have similarities with the periodicity in the apparent viscosity vs shear strain\cite{D.F.Gu}. 
			      The periodicity of this oscillation cycle corresponds to a director rotation angle of $\pi $ at low shear limits\cite{G.Marrucci}. 
			    \\ The VACF of these  structures in all the regimes of nematic phase, can be understood in the context of descriptions of C.R.Safinya \cite{C. R. Safinya}and K.Negita et.al.,\cite{K. Negita}.  From figure\ref{figure 2(b)}, in  $a_{m}$ regime, at $\Delta T_{N-I} = -2.5^{0}C$, the characteristics of precessional motion starts dominating and periodic structures are very clear in this regime. 
			     In $ a_{m}$ regime, amplitude of first harmonic of $C_{v\bot} > C_{v\|}$.   
			      That is, there exists anisotropic precession with larger amplitude in the y-direction. 
			    From figure\ref{figure 2(c)}, in $a_{s}$ regime, the amplitude of first harmonic of $C_{v\|} \simeq C_{v\bot}$. 
			    That is, $ a_{s}$ structure has isotropic precession with equal amplitude in the y and z- directions. 
			    From figure \ref{figure 2(d)}, the amplitude of first harmonic of  $C_{v\|} > C_{v\bot}$. This represents a change in the direction of precessional motion from y to z direction.
			     That is, there exists anisotropic precession with larger amplitude in the z-direction. 
			    From figure \ref{figure 2(f)}, the amplitude of first harmonic of  $C_{v\|} \ll C_{v\bot}$. 
			    This is because, the amplitude of the motion largely gets deflected along the z-axis, which becomes more an out of planed motion of the director, and hence the amplitude of oscillations decreases rapidly. 
			\\
			
			\begin{table}
			\begin{tabular}{|c|c|c|}
			\hline
			Structures&C$_{v\|}$ & C$_{v\bot}$\\
			\hline
			a$_{m}$&0.32&0.42\\
			\hline
			a$_{s}$&0.41&0.35\\
			\hline
			a(b)&0.0054&0.0095\\
			\hline
			a$_{c}$&0.032&0.01\\
			\hline
			
			\end{tabular}
			\caption{Amplitudes of first harmonics of $C_{v\|}$ and $C_{v\bot}$} are multiples of $10^{-15}\frac{m^{2}}{S^{2}}$
			\label{Oscillation amplitude values}
			\end{table}
			
			Further,in order to characterize these structures,  here we present a more qualitative approach  by fitting these VACF with solutions of time dependent harmonic oscillations.
\begin{itemize}
\item $b^{'}$ regime:
			At $ \Delta T_{N-I} = -0.5^{\circ}C$, both  $C_{v\|}$ and  $C_{v\bot}$ fits with solutions of equation of motion(EOM) of under damped harmonic oscillations. Hence, 
			\begin{equation}
			C_{v\|,\bot} = C_{0} e^{\frac{-t}{2\tau_{p}}}  \left( cos \omega t + \frac{sin \omega t}{2 \omega \tau_{p}}\right) 
			\label{underdamped}
			\end{equation}
 \item $a_{m}$ regime:
			At  $ \Delta T_{N-I} = -2.5^{\circ}C$, $Cv\|$ fits with solutions of EOM of coupled harmonic oscillations. Hence,
			\begin{equation}
			C_{v_{\|}} = C_{0} \frac{\omega}{2 \pi \tau_{p}}  \left( \frac{sin(\omega t)}{\omega t}\right) 
			\label{coupled harmonic oscillator}
			\end{equation}
			In this case, $C_{v\bot}$ fits with solutions of decaying coupled harmonic oscillations.
			\begin{equation}
			C_{v\bot} = C_{0} \frac{\omega}{2 \pi \tau_{p}}  e^{\frac{-t}{2\tau_{p}}} \left( \frac{sin(\omega t)}{\omega t}\right) 
			\label{coupled harmonic oscillator with decay}
			\end{equation}
			 
		\item	 $a_{s}$ regime:
		From  $ \Delta T_{N-I} = -3.5^{\circ}C$ to  $ \Delta T_{N-I} = -4.5^{\circ}C$, both  $C_{v\|}$ and  $C_{v\bot}$ fits with solutions of EOM of decaying coupled harmonic oscillations.
		
		   \item $a(b)$ regime:
		    At $ \Delta T_{N-I} = -5.5^{\circ}C$, both  $C_{v\|}$ and  $C_{v\bot}$ fits with solutions of equation of motion(EOM) of half critically damped harmonic oscillations.
		    \begin{equation}
		    C_{v\|,\bot} = C_{0} e^{\frac{-t}{4\tau_{p}}}  \left( cos \omega t\right) 
		    \label{half critically damped oscillations}
		    \end{equation}
		    \item $a_{c}$ regime:
		    At $ \Delta T_{N-I} = -6.5^{\circ}C$, both  $C_{v\|}$ and  $C_{v\bot}$ fits with solutions of equation of motion(EOM) of critically damped harmonic oscillations.
			\begin{equation}
			C_{v\|,\bot} = C_{0} e^{\frac{-t}{4\tau_{p}}} \left[ C_{1} + (C_{2} x)\right] 
			\label{critically damped oscillations}
			\end{equation}
			\end{itemize}
			In these various forms of harmonic oscillator equations, $ C_{0}$ is the normalized values of $C_{v}$, $\tau_{p}$ represents relaxation time of  $\hat{\textbf{n}}$, $\omega$ is the frequency of oscillation of $\hat{\textbf{n}}$, and $C_{1}$, $C_{2}$ are constants. The orienatational state of  $\hat{\textbf{n}}$ can be determined by viscous and elastic torque forces acting on it. In $b^{'}$ regime, since  $(\hat{\textbf{n}})$ makes a small angle with  $\hat{\textbf{V}}$ or  $\hat{\textbf{b}}$ axis, so both $C_{v\|}$ and $C_{v\bot}$ are similar to those measured by T.Turiv et.al.,\cite{T. Turiv} for 5CB liquid crystals.  Hence, they fit well with underdamped harmonic oscillator solutions. In  $a_{m}$ regime, there is no damping term in the equation \ref{coupled harmonic oscillator} of $C_{v\|} $ which represents coupled harmonic oscillator motion.  But, there is a damping term in the equation\ref{coupled harmonic oscillator with decay} of $C_{v\bot}$. This represents beginning of formation of structures with in this regime and is marginally stable \cite{C R Safinya-1}. From  synchrotron X-ray studies,  $a_{s}$ regime is known to be stable with undamped precessing motion. But, from equation \ref{coupled harmonic oscillator with decay}, which represents both $C_{v\|}$ and $C_{v\bot}$ have damping term. Though, damping term is present, but its value is relatively less in this regime. Because of anisotropic suppression of Sm-A fluctuations due to flow, $a(b)$ regime coexists with regions of the sample where $(\hat{\textbf{n}})$ points along X-axis and regions where $(\hat{\textbf{n}})$ points along Y-axis. Hence $C_{v\|},C_{v\bot}$ fits with half critically damped oscillatory equations. In $a_{c}$ regime, from \cite{C R Safinya-1}, $(\hat{\textbf{n}})$ points mostly along X-axis but, tumbles towards Z-axis. But, considering our sample plane, $(\hat{\textbf{c}})$ is along Z-axis, hence we could not measure any precessional motion along this direction.
			\begin{figure}
\includegraphics[scale=0.35]{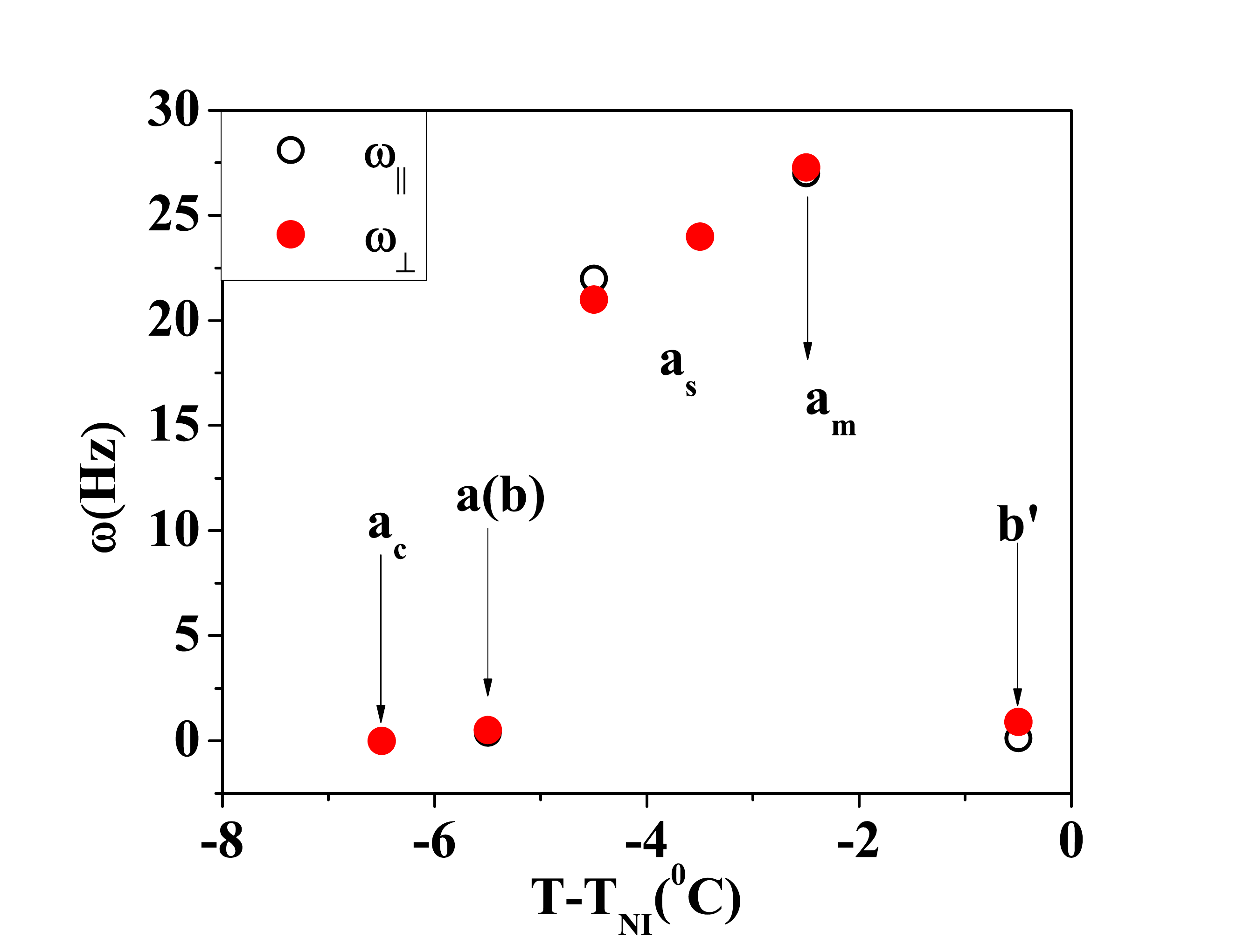}

\caption{angular frequency $ (\omega) $ vs $ \Delta T $ of precessional motion of $(\hat{\textbf{n}})$  in different regimes of nematic phase of 8CB liquid crystals. $(\omega_{\|})$ is shown by open circles and $(\omega_{\|})$ is shown by closed (red) circles.}
\label{figure 3}
\end{figure}

\begin{figure}
\includegraphics[scale=0.35]{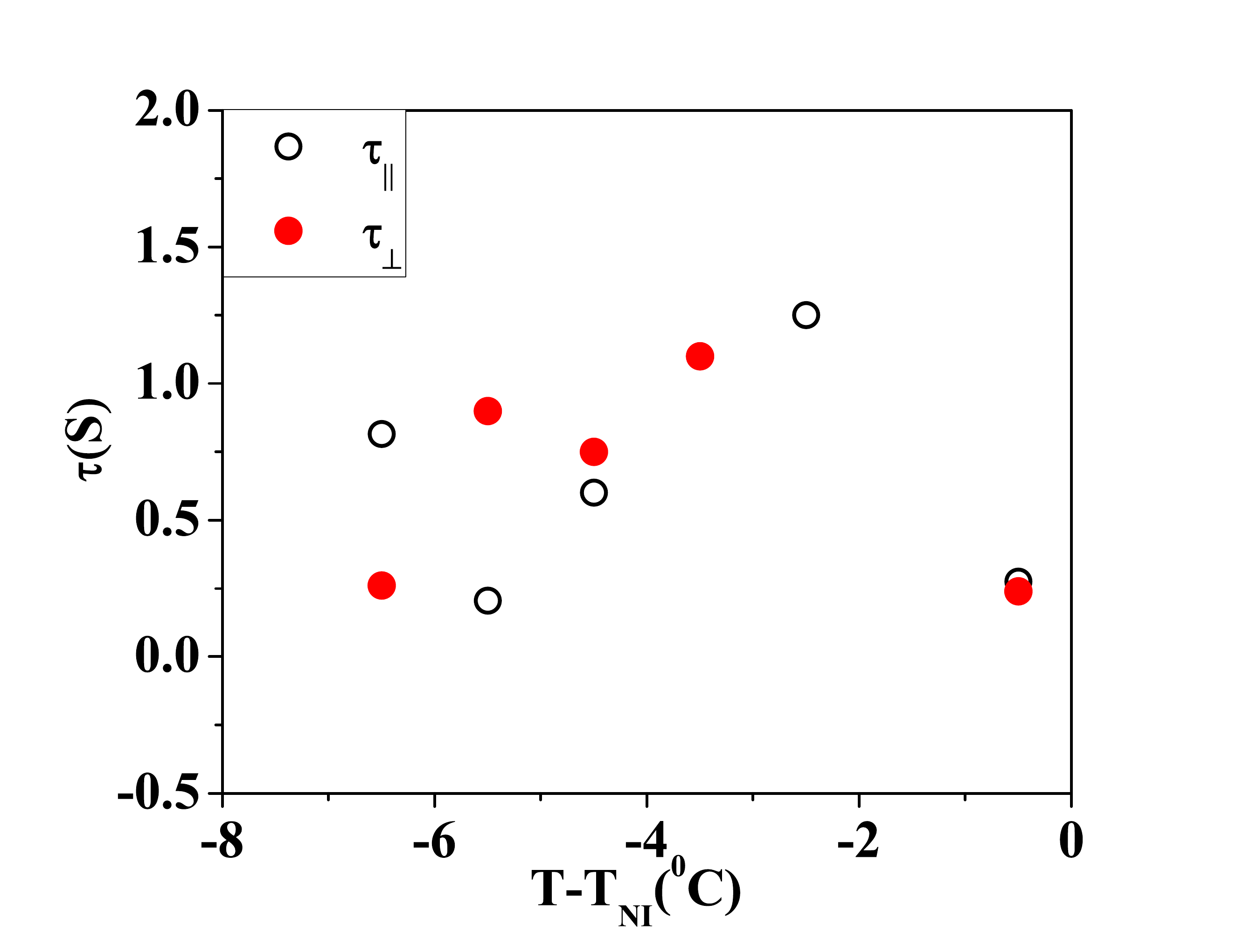}

\caption{Time of relaxation $(\tau) $ vs $ \Delta T $ of $(\hat{\textbf{n}})$  in different regimes of nematic phase of 8CB liquid crystals. $(\tau_{\|})$ is shown by open circles and $(\tau_{\|})$ is shown by closed (red) circles.}
\label{figure 4}
\end{figure}

			Also, microstructural properties like  periodic oscillation frequency$(\omega)$, relaxation time$(\tau)$ of $\hat{\textbf{n}}$ are quantified by fitting $C_{v\|}$ and $C_{v\bot}$ with harmonic oscillator equations. From these properties, micro mechanical properties like elastic constant $(K)$(one elastic constant approximation), quality factor(Q) that describes stability of the precessional motion of $\hat{\textbf{n}}$ are calculated. These results predicts the similar nature to the precessional motion of $\hat{\textbf{n}}$  along different axes $V, \Delta V $ and neutral axis as mentioned in  previous experimental results\cite{C. R. Safinya},\cite{K. Negita-1} which are based on Ericksen-Leslie-Parodi(ELP) theory and critical nemato dynamics. In figure\ref{figure 3}, for both $C_{v\|}$ and $C_{v\bot}$, precessional frequency increases from regimes $b^{'}$ to $a_{m}$ because of pretransitional Sm-A fluctuations and then it decreases rapidly as it reaches $a(b)$ regime and is zero in $a_{c}$ regime, where, pretransitional Sm-A fluctuations is expected to give rise to flow induced fluctation force similar to $b'$ regime. Further, the measured values of $\tau$ represent the slow dynamics and are inversely proportional to viscosity$(\sim \frac{1}{\gamma})$. In figure\ref{figure 4}, from $b'$ to $a_{m}$ regime, $\tau$ increases, and it decreases almost steeply till $a(b)$. In $a_{m}$ regime, at $ \Delta T_{N-I} = -3.5^{\circ}C$, since $C_{v\bot}$ fits with coupled harmonic oscillator equation, with no damping term, the value of $\tau$ is not presented. This $\tau$ vs $\Delta T$ shows inversely proportionate behavior with  $\eta$ vs $\Delta T $ in shear rheology experimental results\cite{K. Negita-1}. The decrease in the value of $\tau_{\|}$ at $ \Delta T_{N-I} = -5.5^{\circ}C$, in $a(b)$ regime, can be attributed to  the pretransitional Sm-A fluctuations which can be expected to give rise to a flow induced fluctuation force\cite{C. R. Safinya}.  But, there is a crossover in $\tau_{\|}$ and $\tau_{\bot}$ values in $a_{c}$ regime $ \Delta T_{N-I} = -6.5^{\circ}C$. Here, the  liquid crystal cell is rubbed, so the smectic layer planes in the temporarily fluctuating clusters, that are oriented perpendicular to the rubbing direction can be expected to have slow relaxation times\cite{S. Dhara}. 
 Therefore, $\tau_{\bot}$ is lower than $\tau_{\|}$.
 \begin{figure}
\includegraphics[scale=0.35]{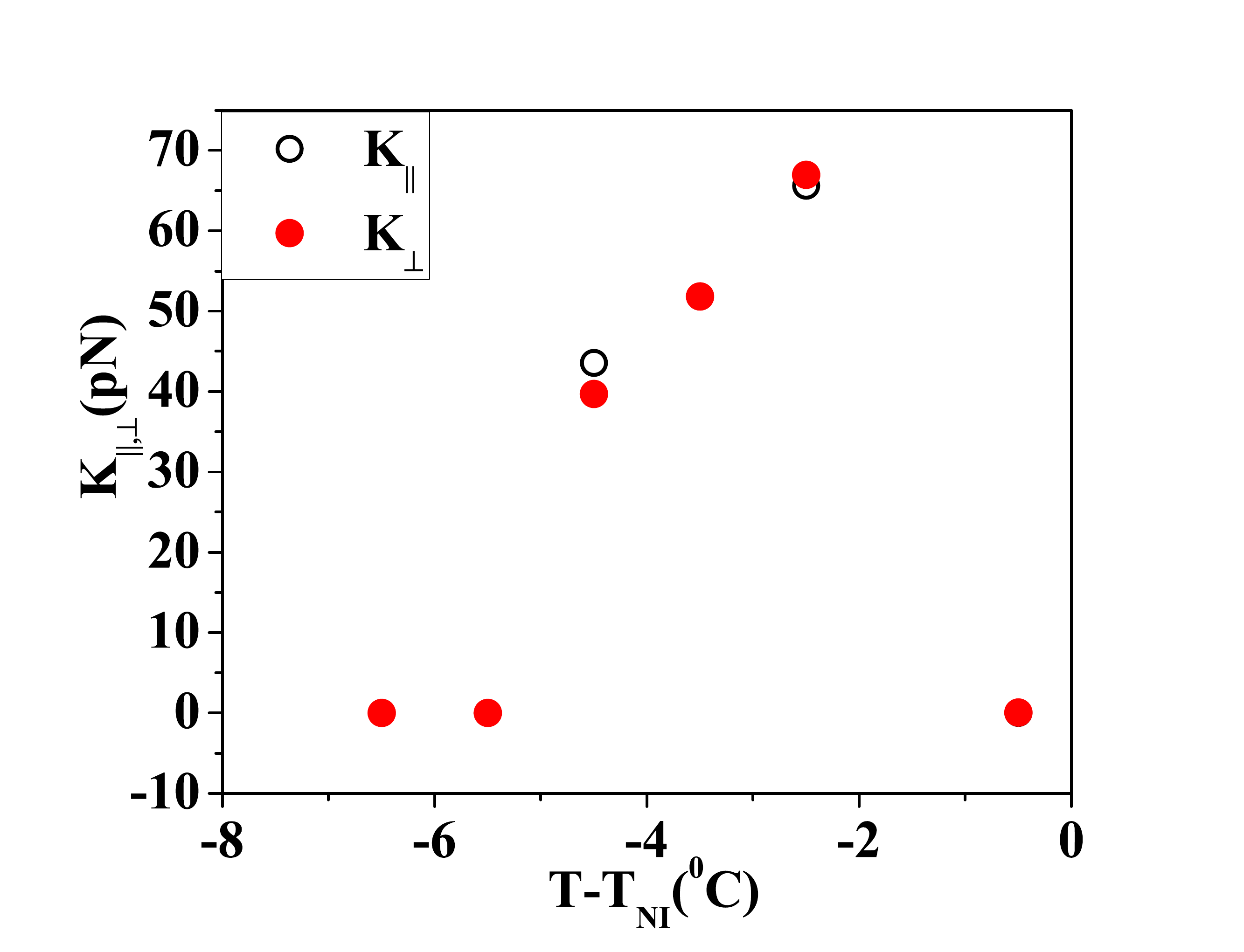}
\caption{Elastic constant $(K)$ vs $ \Delta T $ in different regimes of nematic phase of 8CB liquid crystals. $(K{\|})$ is shown by open circles and $(K_{\|})$ is shown by closed (red) circles.}
\label{figure 5}
\end{figure}

\begin{figure}
\includegraphics[scale=0.35]{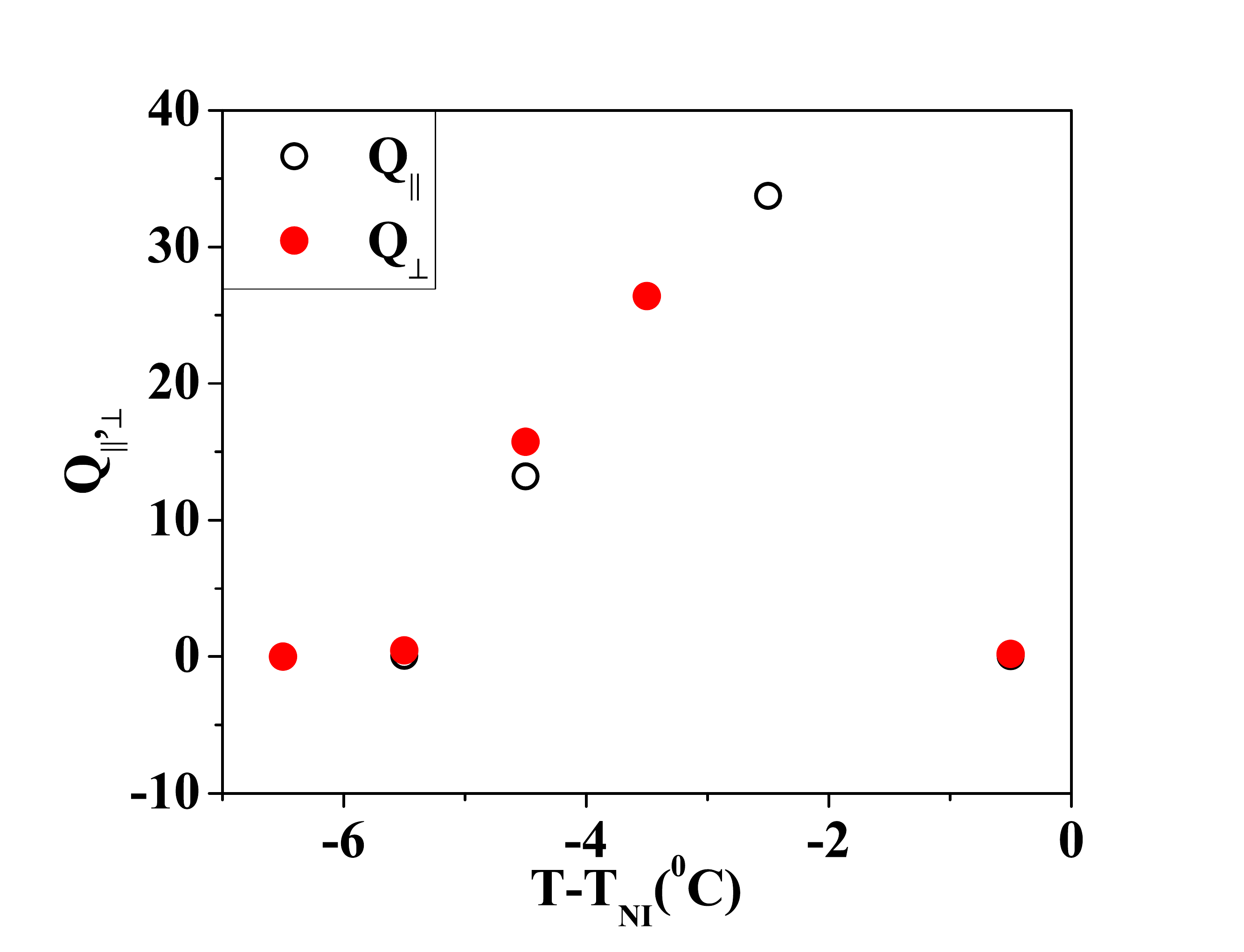}

\caption{Quality factor $(Q)$ vs $ \Delta T $ of precessional motion of  $(\hat{\textbf{n}})$  in different regimes of nematic phase of 8CB liquid crystals. $(Q_{\|})$ is shown by open circles and $(Q_{\|})$ is shown by closed (red) circles.}
\label{figure 6}
\end{figure}

  Negative values of auto correlations in $C_{v}$ indicates a tendency to move back to earlier position. This antipersistant behavior is the characteristic of viscoelastic medium.  From the measured values of $\omega$, elastic constant $(K)$ can be calculated $(K = m\omega^{2})$. Obviously, the characteristics of K replicate the same characteristics of $\omega$. In $a_{m}$, $a_{s}$ regimes elastic nature is more dominant as shown in figure\ref{figure 5}. Further, in case of $\frac{1}{\tau}\ll\omega$, the quality factor$(Q)$ of harmonic oscillations is $Q = \omega \tau$. This $Q$ represents the stability of the harmonic oscillations. From figure\ref{figure 6}, it is clear that, both $a_{m}$ and $a_{s}$ are more stable regimes.
\maketitle
\section{Conclusions}
	The challenge of next step to identifying anomalous behavior lies in determining the underlying mechanism.  As the colloidal particle passes through 8CB liquid crystals medium, the precessing nematic director ``pushes back'' creating long time correlations in the particle trajectory. This memory leads to sub diffusive behavior. In any disordered or complex medium, colloidal particle gets navigated through disordered structures, when the colloidal probe is of the same diameter as of the wavelength of disordered structures.
 For the first time, single particle tracking technique is used in determining the nature of the amplitudes of  precessional motions of different structures in 8CB liquid crystals. Advantage of this technique is in tracking these precessional motions in parallel and perpendicular directions to the nematic director. VACF $(C_{v})$, which describes the intrinsic memory of the dynamical structures is used in determining diffusive behavior of the colloidal particle. The $C_{v\|}$ and  $C_{v\bot}$ have shown the profound nature of the precessional motions of the nematic director. The amplitudes of these precessional motions in the parallel direction are very much in agreement with the precessional motion amplitudes measured using bulk rheometer. The amplitudes of precessional motions of the nematic director in the perpendicular direction was measured for the first time and observed a continuous decrease in these peak amplitudes. Also, the amplitudes of isotropic and anisotropic precessional motions in various structures with in the nematic phase were observed and they are in well agreement with the previous literature.  Further, we have measured microstructural and micromechanical properties of the 8CB liquid crystals in its nematic phase by fitting the oscillatory motion present in the VACF of 8CB liquid crystals with the time dependent harmonic oscillations. This fitting with time dependent harmonic oscillator motion resulted in identifying the transitions between different structural regimes in the neamtic phase. Futher, an important aspect of this study identifies a difference between shear rheology measurements and zero shear conditions. As per ELP theory, it is true that, in these complex fluids this dynamics of $\hat{\textbf{n}}$ arises because of the shear forces which applies torque to the liquid crystal molecules. But, as per the understanding of shear rheology experiments, that shear forces need not be external.   To our knowledge,  an attempt using Brownian motion of the colloidal sphere in the tumbling nematics is attempted for the first time. We believe that, this is a necessary and successive step after the experimental measurements by T.Turiv et.al.,\cite{T. Turiv}.  Many complex fluids  like actomysin cells \cite{Sundar}, lipid bilayers \cite{Gueguen} follow the dynamics observed in this study with 8CB liquid crystals. Hence, this generic approach of analyzing with time dependent harmonic oscillations can be applied to those biological systems. As a future direction, understanding the relation between normalization constants $(C_{0})$ in the above mentioned time dependent harmonic  oscillator equations and Leslie coefficients should be established.
 
%

\end{document}